\documentclass[sigconf]{acmart}
\usepackage{todonotes}
\usepackage{subfiles} % Best loaded last in the preamble
\usepackage{listings, color}
\usepackage{makecell}
\usepackage[framemethod=default]{mdframed}
\usepackage{amsthm}
\usepackage{tabularx}
\usepackage{hhline}
\usepackage{graphicx}
\usepackage{caption}
\usepackage{subcaption}
\usepackage{array, booktabs, makecell, multirow}
\usepackage{mathtools}
\setcellgapes{5pt}
\definecolor{viatraEmphColor}{RGB}{0,80,125}
\definecolor{keywordcolor}{rgb}{0.5,0,0.1}
\definecolor{commentcolor}{rgb}{0,0.3,0.1}
\definecolor{stringcolor}{rgb}{0,0,1}

\lstdefinelanguage{viatra}
{
morekeywords={@QueryBasedFeature,@Constraint,count,pattern,package,neg,find,import,true,false,or,check,job,action,state,severity,location,message,oclIsKindOf,self,exists,includes,invariant,class,private,epackage,java},
emph={Pseudostate,Vertex,Region,Transition,Entry,Synchronization,State,RegularState,CompositeElement,Trigger,Guard,Action,Statechart,vertices,regions,source,target,incomingTransitions,trigger,guard,action,elements,newElements,PartialModel,open,must,may,var,element,transitions,FamilyTree,members,Member,parents,age,VisionBlocked,blockedBy,Actor,xPos,yPos,ySpeed,placedOn,Lane_Horizontal,length,width},
emphstyle={\color{viatraEmphColor}},
sensitive=true, morecomment=[l]{//}, morecomment=[s]{/*}{*/},
morestring=[b]{"}
}
\lstdefinestyle{viatrasmall}{
	basicstyle=\scriptsize\ttfamily,
	commentstyle=\color{commentcolor}\ttfamily,
	stringstyle=\color{stringcolor}\ttfamily,
	captionpos=b,
	keywordstyle=\color{keywordcolor}\bfseries\ttfamily,
	showstringspaces=false,
	tabsize=2,
	language=viatra,
	escapeinside={(*@}{@*)}%,
}
\lstdefinestyle{viatrabig}{
	basicstyle=\ttfamily,
	commentstyle=\color{commentcolor}\ttfamily,
	stringstyle=\color{stringcolor}\ttfamily,
	captionpos=b,
	keywordstyle=\color{keywordcolor}\bfseries\ttfamily,
	showstringspaces=false,
	tabsize=2,
	language=viatra,
	escapeinside={(*@}{@*)}%,
}

\definecolor{emphColor}{rgb}{0.1,0.1,0.1}  %grey

\definecolor{logsemColor}{RGB}{0,80,125}

\definecolor{numsemColor}{RGB}{204,100,0}

\newcommand{\ranswer}[2]{%
\medskip
\noindent
\begin{tabularx}{\linewidth}{|X|}\hline
\rquestion{#1}: \emph{#2}\\\hline
\end{tabularx}\medskip}

\usepackage{xspace}

\definecolor{blue0}{HTML}{47BCFF}
\definecolor{blue1}{HTML}{009BF5}
\definecolor{blue2}{HTML}{0067A3}
\definecolor{blue3}{HTML}{003452}
\definecolor{car1}{RGB}{128,125,123}  %grey
\definecolor{car2}{RGB}{194,92,85}  %red
\definecolor{car3}{RGB}{75,119,157}  %blue

\newtheorem{example}{Example}

\newcommand{\rquestion}[1]{\textbf{\textsc{RQ}#1}}
\definecolor{completed}{RGB}{51,153,102}

\definecolor{inProgress}{RGB}{196,196,35}

\definecolor{notYetStarted}{RGB}{255,0,0}

\NewDocumentCommand{\rot}{O{45} O{1em} m}{\makebox[#2][l]{\rotatebox{#1}{#3}}}%
\newcommand{\halfcheck}{X\kern-1.1ex\raisebox{.7ex}{\rotatebox[origin=c]{125}{--}}}

\newcommand{\kg}{KG}
\newcommand{\ir}{IR}
\newcommand{\mis}{MIS}

\DeclareMathOperator*{\argmax}{arg\,max}
\AtBeginDocument{%
  \providecommand\BibTeX{{%
    \normalfont B\kern-0.5em{\scshape i\kern-0.25em b}\kern-0.8em\TeX}}}

\copyrightyear{2022}
\setcopyright{acmcopyright}
\acmYear{2022}
\acmConference[TrustLOG-CIKM '22]{The First Workshop on Trustworthy Learning on Graphs in Conjunction with the 31st ACM International Conference on Information and Knowledge Management}{October 21, 2022}{Atlanta, GA, USA}
\acmBooktitle{The First Workshop on Trustworthy Learning on Graphs in Conjunction with the 31st ACM International Conference on Information and Knowledge Management (TrustLOG-CIKM '22), October 21, 2022, Atlanta, GA, USA} 
\acmPrice{15.00}

\begin{document}

%%
%% The "title" command has an optional parameter,
%% allowing the author to define a "short title" to be used in page headers.
\title[Towards Improving the Explainability of Text-based Information Retrieval with Knowledge Graphs]{Towards Improving the Explainability of \\ Text-based Information Retrieval with Knowledge Graphs}
% Knowledge graph, (embedding) information retrieval, improvement, trustworthy (explainability)

%%
%% The "author" command and its associated commands are used to define
%% the authors and their affiliations.
%% Of note is the shared affiliation of the first two authors, and the
%% "authornote" and "authornotemark" commands
%% used to denote shared contribution to the research.
\author{Boqi Chen}
\authornote{All three authors contributed equally to this research.}
\affiliation{%
  \institution{McGill University}
  \city{Montreal}
  \state{Quebec}
  \country{Canada}
}
\email{boqi.chen@mail.mcgill.ca}

\author{Kua Chen}
\authornotemark[1]
\affiliation{%
  \institution{McGill University}
  \city{Montreal}
  \state{Quebec}
  \country{Canada}
}
\email{kua.chen@mail.mcgill.ca}

\author{Yujing Yang}
\authornotemark[1]
\affiliation{%
  \institution{McGill University}
  \city{Montreal}
  \state{Quebec}
  \country{Canada}
}
\email{yujing.yang2@mail.mcgill.ca}

\author{Afshin Amini}
\affiliation{%
  \institution{Aggregate Intellect}
  \city{Toronto}
  \state{Ontario}
  \country{Canada}
}
\email{afshin.emd@gmail.com	}

%% (10/11) Bharat - i updated the city and state information.
\author{Bharat Saxena}
\affiliation{%
  \institution{BMC Software Inc}
  \city{Pune}
  \state{MH}
  \country{India}
}
\email{bharat_saxena@bmc.com	}

\author{Cecilia Chávez-García}
\affiliation{%
  \institution{Aggregate Intellect	}
  \city{London}
  \state{Ontario}
  \country{Canada}
}
\email{cecilia.uku@gmail.com}

\author{Majid Babaei}
\affiliation{%
  \institution{McGill University}
  \city{Montreal}
  \state{Quebec}
  \country{Canada}
}
\email{majid.babaei@mcgill.ca	}

\author{Amir Feizpour}
\affiliation{%
  \institution{Aggregate Intellect}
  \city{Montreal}
  \state{Quebec}
  \country{Canada}
}
\email{amir.fzpr@gmail.com	}

\author{D\'aniel Varr\'o}
\affiliation{%
  \institution{McGill University}
  \city{Montreal}
  \state{Quebec}
  \country{Canada}
}
\email{daniel.varro@mcgil.ca}

%%
%% By default, the full list of authors will be used in the page
%% headers. Often, this list is too long, and will overlap
%% other information printed in the page headers. This command allows
%% the author to define a more concise list
%% of authors' names for this purpose.
\renewcommand{\shortauthors}{B. Chen, K. Chen, Y. Yang, A. Amini, B. Saxena, C. Chávez-García, M. Babaei, A. Feizpour and D. Varr\'o}

%%
%% The abstract is a short summary of the work to be presented in the
%% article.
\begin{abstract}
Thanks to recent advancements in machine learning, vector-based methods have been adopted in many modern information retrieval (\ir{}) systems. While showing promising retrieval performance, these approaches typically fail to explain why a particular document is retrieved as a query result to address explainable information retrieval (XIR). Knowledge graphs record structured information about entities and inherently explainable relationships. 
% Existing XIR approaches focus exclusively on the retrieval model without considering existing knowledge graphs for providing an explanation. 
Most of existing XIR approaches focus exclusively on the retrieval model with little consideration on using existing knowledge graphs for providing an explanation.
In this paper, we propose a general architecture to incorporate knowledge graphs for XIR in various steps of the retrieval process. Furthermore, we create two instances of the architecture for different types of explanation. We evaluate our approaches on well-known \ir{} benchmarks using standard metrics and compare them with vector-based methods as baselines.
% and their application to a wide range of areas, there is a growing research interest in explainable artificial intelligence (XAI) to improve the interpretability, explainability, and transparency of the system.
\end{abstract}

%%
%% The code below is generated by the tool at http://dl.acm.org/ccs.cfm.
%% Please copy and paste the code instead of the example below.
%%


% \ccsdesc[500]{Computer systems organization~Embedded systems}
% \ccsdesc[300]{Computer systems organization~Redundancy}
% \ccsdesc{Computer systems organization~Robotics}
% \ccsdesc[100]{Networks~Network reliability}

%%
%% Keywords. The author(s) should pick words that accurately describe
%% the work being presented. Separate the keywords with commas.

%% A "teaser" image appears between the author and affiliation
%% information and the body of the document, and typically spans the
%% page.

%%
%% This command processes the author and affiliation and title
%% information and builds the first part of the formatted document.
\maketitle

\section{Introduction}
\label{intro}

% With the help of Internet, people now could have easy and convenient access to massive amount of information. For example, using Google search has become an essential part of people's work and life. Information retrieval is defined as the process of accessing the most relevant information with the input user query. This research specifically focuses on document retrieval defined as: given a query q and a collection of document D, score and rank each document d belongs to D based on its relevance to q \cite{p5}.
% The problem in the field is that Traditional embedding based retrievers rank document based on cosine similarity or dot product on vectors and they presents little explainability to users\cite{p3}. Our goal is to make retrieval systems more explainable and better answer the question: why a document is selected and why a document is ranked at a certain position. This field is called explainable information retrieval (XIR) and is a subcategory of explainable AI (XAI).

\paragraph{Motivation}
Given a query in natural language as input, text-based information retrieval (\ir{}) is a task of accessing the most relevant documents. Modern \ir{} systems usually exploit text embeddings, which map text into a vector such that the distance between the vectors represents the similarity of the text. 

In real-world \ir{} use cases, a user can make better decisions if additional explanation is provided to the retrieved results. For example, when a user queries what contributes to heart disease, an explanation stating the reason help the user decide the relevance of a result. This is \textit{local} explanation which addresses why a document $d$ is relevant to the user query $q$. In general, an \textit{local} explanation should be (1) \textit{interpretable} i.e. it should be understandable by a human and (2) \textit{accurate} i.e. it should accurately answer the question.

Common explanations may highlight the most relevant parts of a document \cite{p2} or provide feature importance score \cite{Ramos2019ExplainabilityIT}. Recent approaches also use explainable features to re-rank candidates generated by vector-based \ir{} systems \cite{polley2021exdocs}. However, few existing approaches explicitly incorporate the structured information from the text, which may serve as an important source for explanations.

% A knowledge graph (\kg{}) is a directed labelled graph in which domain-specific meaning are associated with nodes and edges. The edges capture relationships between node representing entities. The recorded information is normally both understandable by human and machine \cite{chaudhri2022knowledge}, which means a \kg{} is inherently explainable. \kg{}s have been used in many different fields to record information in structured manner \cite{abu2021domain}, which means one can normally find an existing \kg{} for the domain of interest. \kg{}s has been shown effective for improving the performance of \ir{} systems \cite{p5} and generating explanation for other machine learning models \cite{tiddi2022knowledge}. However, \kg{} for explainable information retrieval is still restricted to domain-specific use cases \cite{yang2020biomedical, abu2022transferrable}. And how \kg{}s can be used for general text-based \ir{} is yet to be explored. 

\paragraph{Problem statement}
In this paper, we focus on the task of \emph{explainable information retrieval} (XIR), which aims to increase the explainability of retrieval algorithms, i.e., make the result more understandable and reflect the internal behavior of the algorithm. We aim to investigate how knowledge graphs as external sources with explicit semantic relationships can be utilized to improve XIR.  

A knowledge graph (\kg{}) is a directed labeled graph where nodes represent entities, and edges capture relationships between nodes.  \kg{}s have been shown to be effective for improving the performance of \ir{} systems \cite{p5} and generating explanation for other machine learning algorithms \cite{tiddi2022knowledge}. However, the use of \kg{} for explainable \ir{} is still restricted to domain-specific use cases \cite{yang2020biomedical, abu2022transferrable}. 

\kg{}s provide explicit semantic relations between entities, which can serve as  explanation, or help clarify the intent of the query. Normally, such semantic relations also appear in the text. For example, the sentence "obesity contributes to heart disease" reflects two entities \textit{obesity} and \textit{heart disease} related by the \textit{contribute} relation. 

Linking entities and relationships, i.e. identifying and connecting the matching parts of a \kg{} and a piece of text, is a well-known  challenge \cite{tagme, wat}. While existing entity linking tools can be used as a starting point, improving the explainability of \ir{} by using \kg{}s also necessitates (1) a general integrated architecture, and (2) a detailed recipe on how to use \kg{} information to improve specific types of explanations.

\paragraph{Objective and contribution}
% Our approach use knowledge graph to improve the explainablity of text-based information retrieval in two aspect: highlight the most important sentence and re-rank documents with explainable matrix.
Given a set of natural language documents and a query, our paper proposes a general architecture to integrate \kg{} into \textit{vector-based} \ir{} systems to generate \textit{local} explanation for the retrieved results. We use various entity/relationship linking methods to match the most relevant parts of the \kg{} to the text and then use this information to improve the explainability of the \ir{} system. In summary, the specific contributions are:
\begin{itemize}
    \item We propose a general architecture that divides \ir{} into different phases and highlights how \kg{} can be exploited for better explanation in each phase. 
    \item We instantiate the general architecture in two ways for explainable IR by integrating an open knowledge graph Wikidata and various entity linking approaches.
    \item We conduct initial experimental evaluations on the two specific explanation approaches on two datasets: WIKIQA \cite{YangYihMeek:EMNLP2015:WikiQA} and Robust04 \cite{robust04}. Assuming an ideal entity matcher, our technique improves the performance of identifying the most important sentence by 6.96\% and the base retrieval performance by 0.45\%.
\end{itemize}

\begin{figure}[tb]
    \centering
\includegraphics[width=0.8\columnwidth]{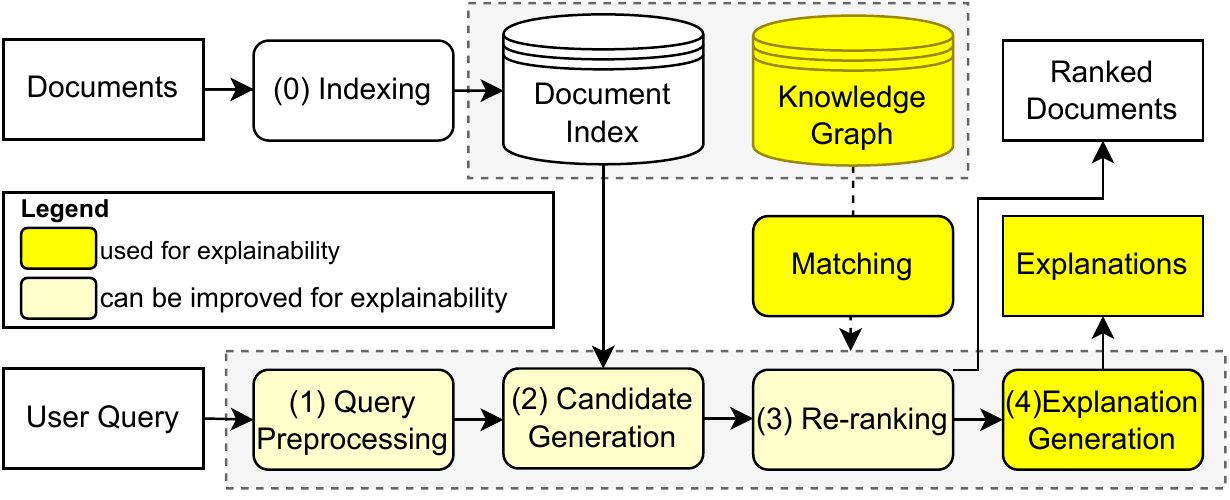}
\caption{General architecture for explanation with KG}
    \label{fig:architecture}
\end{figure}
%\end{figure*}

\section{Approach}
\label{approach}

\subsection{General Architecture}
\autoref{fig:architecture} shows the general architecture of our explainable \ir{} system enhanced with \kg{}s. In a preprocessing phase, (0) documents are stored and arranged into the \emph{document index} (assumed to store documents in a vectorized form). Then (1) our framework takes a user query in natural language as input, and \emph{preprocesses} the query (e.g. by query expansion \cite{azad2019query}). Next (2) \emph{candidate generation} retrieves a list of ranked candidate documents which are most relevant to the processed query. (3) This list of candidate documents may be rearranged by \emph{re-ranking}. Finally, (4) our  framework outputs ranked documents together with the output from \textit{explanation generation}. 

Our architecture incorporates how extra knowledge from \kg{} can be exploited using entity/relationship \textbf{matching} in each step (1)-(4) of \ir{} to gradually improve explainability.
%provided finally in the \emph{Explanation Generation} phase. 

%integrates the improvements from each individual step to The improvement from \kg{} flows from parts to parts and finally helps to improve the quality of the .

\begin{itemize}
    \item \textbf{Entity/relation matching:} 
    %The entity can be identified from the query and linked to the external knowledge base with \textit{matching}. 
    % In order to utilize a \kg{} for an \ir{} task, one normally needs to perform a linking task from the text to the \kg{}. The task includes entity recognition, relationship extraction, and entity linking with \kg{}. 
    Bridging text with corresponding entities in a knowledge base is carried out by entity/relation matching \cite{shen2014entity}. This step first uses entity recognition to detect keywords and entities in the query, such as proper names \cite{mansouri2008named}. These entities are matched to nodes in the \kg{}. Then relation extraction is applied to identify the relations between instances and concepts of corpus data, which transforms identified entities into structured data \cite{lin2015learning}. 
% With an effective entity linking system, the named entity is linked to a knowledge base that is derived from a credible source, and the query is expanded with related features, and representations \cite{dalton2014entity}.
    \item \textbf{Query preprocessing for explainability:}
The user query can be expanded with information that is retrieved from the \kg{}. A \kg{} normally contains both entity information such as description, alias and properties, as well as relation information such as connected entities and relation types. Such extra knowledge can be added to the query to improve query understanding.  
    \item \textbf{Explainable candidate generation:}
    Explainable candidate generation may use a \kg{} constructed directly from  documents themselves. This \kg{} is built from entities and relations identified from the documents where each entity may be related to one or more documents. Then, the candidate generation becomes a \kg{} retrieval problem and it can be performed using exact or fuzzy entity matching.
    
    \item \textbf{Explainable re-ranking:}
    % After \textit{matching}, an explainable feature of query-document relatedness is proposed, which relies on entity relatedness calculated from \kg{} to find how two texts are related to each other. 
    The \kg{} can also be used to calculate explainable features for re-ranking. For example, both the query and candidate document can be first translated into a set of entities. Then pair-to-pair relatedness of the entities can be calculated using their distance and connectivity on the \kg{}. Such score can be aggregated into query-document relatedness to be used for re-ranking. 
    
    %is calculated and aggregated to represent the relatedness of two texts which, in our case, are the query and the document.
    \item \textbf{Explanation generation:}
    % Entities detected from a passage can be an effective way of informing users the content of the passage without looking through it. Another approach to generate explanations could be building a new knowledge graph based on the entities and relations detected from the passage. Basically, a passage can be translated into a knowledge graph so that the information is extracted into a compact structured form.  
    Information from steps (1)-(3) can be used to generate enhanced explanations to the user. The expanded query from (1) can be used to find the most relevant part of a retrieved document and explain why a query is related to a document. \kg{} used in (2) can provide the number of matching entities between the query and documents. Finally, features calculated in (3) provide intrinsically explainable ranking of the documents.    
\end{itemize}

\subsection{\kg{} for Most Important Sentence}
\begin{figure}[tb]
    \centering
\includegraphics[width=0.8\linewidth]{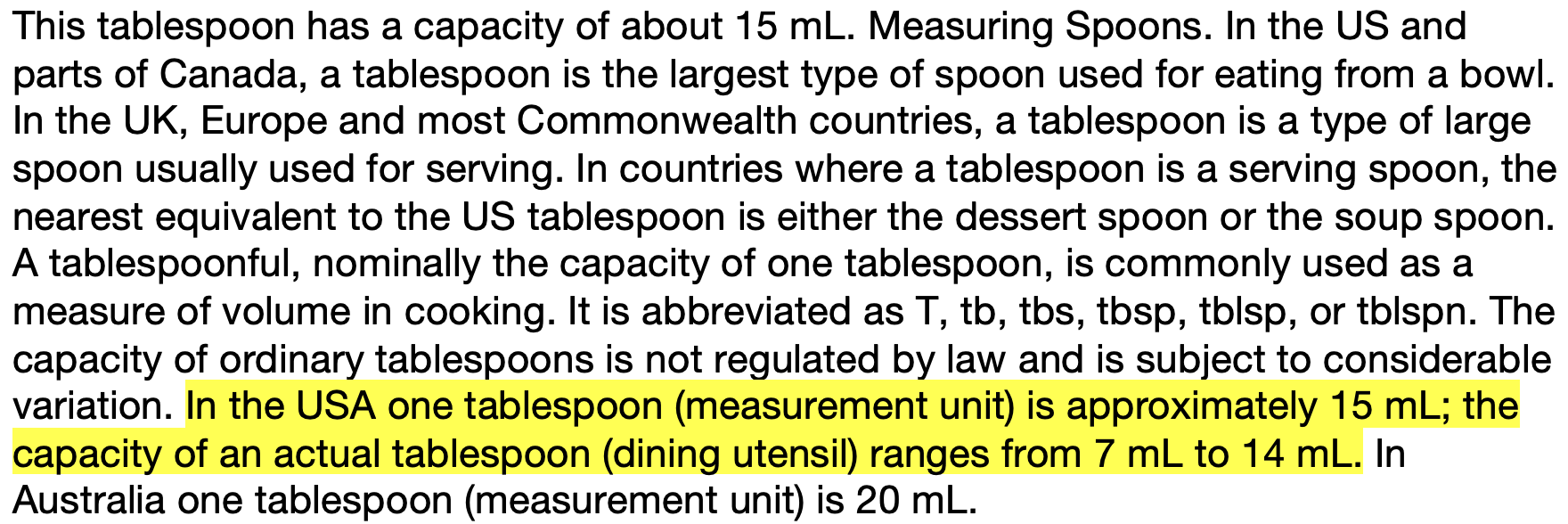}
\caption{Example of \mis{}}
\label{fig:highlight}
\end{figure}

%\subsubsection{Visualization of the most important sentence}
%The first explainable element is highlighting the most relevant sentence within the retrieved passage.

% \paragraph{Rationale}
% Why highlight the most important sentence / passage
% What does the most important sentence mean
The most important sentence (\mis{}) is the most relevant part of a passage and explains how the passage is related to the query. The \mis{} attracts the user's attention and emphasizes the main content that relates to the query so that the user can understand why the passage is retrieved without reading the whole passage. %Even if the retriever does not return the correct passage, the user can know there is something wrong without reading the whole content or wasting time. 
%This can be especially helpful in reading long documents. 
%where the information needed by the user could be at the bottom of the document.
Formally, let $q$ be the query and $S$ be a set of sentences from the document, and $sim(\cdot, \cdot)$ be some similarity function between two pieces of text. The \emph{most important sentence} is defined as
\begin{equation}
    MIS=\argmax_{s \in S} sim(q, s)
    \label{equ:mis}
\end{equation}

A method widely used to measure similarity between two pieces of text is embedding. We adapt this method in our approach by embedding the text into a vector. Then a standard similarity metric between two vectors such as cosine similarity or dot product can be used to approximate the similarity function $sim$ in \autoref{equ:mis}. We chose cosine similarity in our approach. 
% where $v_q$ and $v_s$ represent the embedding of the query and sentence. 

\begin{example}
\normalfont
\autoref{fig:highlight} shows an example of highlighting the MIS in a paragraph. The paragraph is about a tablespoon, and the MIS contains the definition and capacity of a tablespoon.
\end{example}
%\begin{example}
% \includegraphics[scale=0.2]{figures/entity_rela.pdf}

% \paragraph{Retrieval Process}
% Retrieval process from figure 1
% various ways of candidate generation. Do not need sentence retriever here

Naively, one can use a sentence retriever that combine the query embedding and the embedding of each sentence in the passage to identify the \mis{} with cosine distance \cite{singhal2001modern}. Yet, this approach normally misses some semantic information about the query entity. For example, the word 'bank' may mean a financial institute as well as construction on the river. To avoid semantic ambiguity in the query, we expand the query using entity linking and relationship extraction with a reference \kg{}. 

% \todo{Connect the last two paragraphs}
% The retrieval process is the \emph{Candidate Generation}  the \autoref{fig:architecture}. 
% %It includes two steps: passage retrieval and sentence retrieval. 
% Here, we assume the retrieval is based on vector search 
% %and do not differ the candidate generation and re-ranking.
% , and we use candidate generation to generate a list of ranked passages without using re-ranking. However, our approach is applicable to any retrieval process. In the passage retrieval process, the candidate sentences are ranked based on their cosine similarity \cite{singhal2001modern} with the query. 

% Architecture Diagram
%\begin{figure}
%    \centering
%\includegraphics[width=0.7\linewidth]{figures/example.jpg}
%\caption{Example of \mis{}}
%    \label{fig:example}
%\end{figure}

% In the passage retrieval process, the passage with the highest cosine similarity score from each query-passage pair is retrieved as the most relevant passage. 

%In the sentence retrieval process, the most relevant passage is split into sentences and the sentence with the highest cosine similarity score is retrieved, using the same method of passage retrieval. Then, the most relevant sentence is used as explanations to the overall retrieval process. 
% example???

% \paragraph{Explanation with \kg}
% How knowledge graph is used to generate the most important sentence

%This external information might improve query's understanding and the performance of information retrieval. 

\begin{figure}[H]
    \centering
\includegraphics[width=.6\columnwidth]{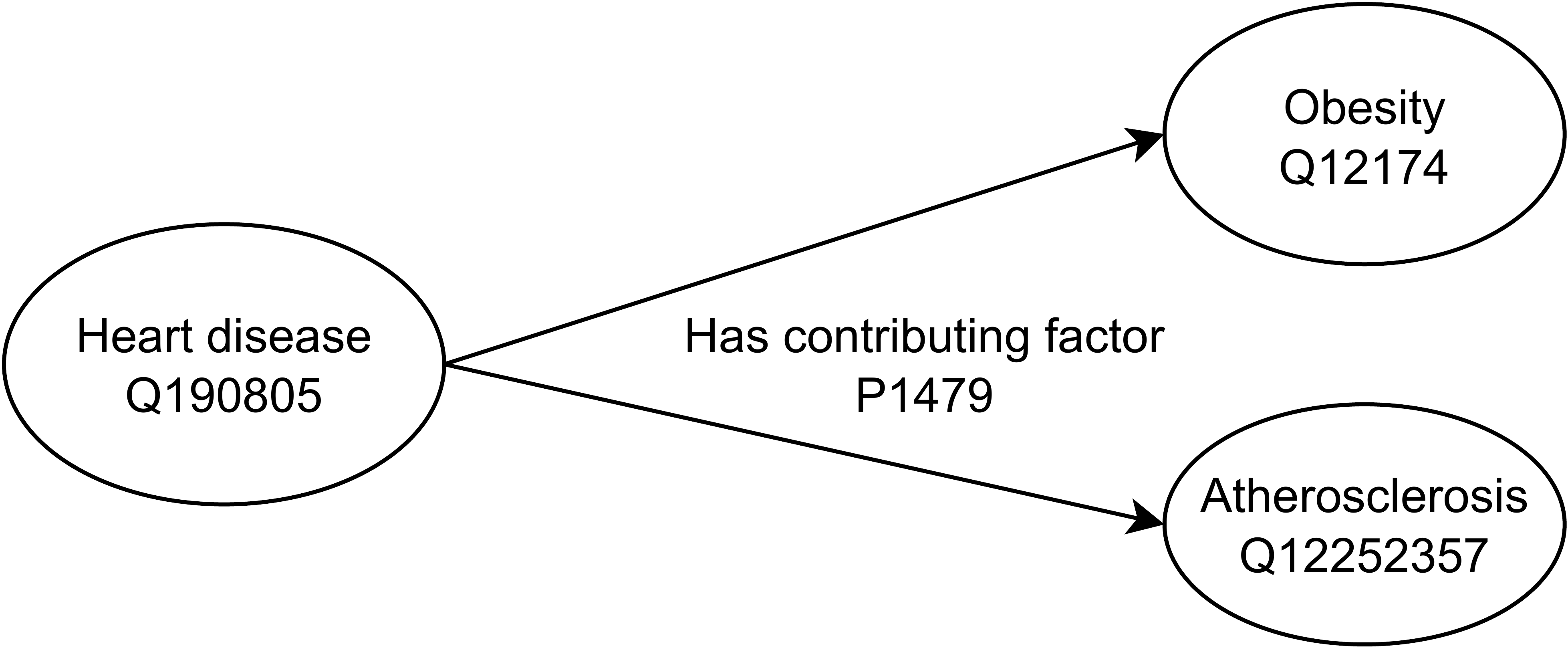}
\caption{Example of entities and relations}
\label{fig:entity_rela}
\end{figure}

We extend \textit{query preprocessing} to better identify \mis{} while keeping other parts of the architecture unchanged. Specifically, we use extra information of the linked entity from the \kg{} to perform \emph{query expansion} in three different ways. 
\begin{enumerate}
    \item[(A)] If the query contains an entity with a direct relationship, the linked entity is considered as potential information, and it is appended to the query (see \autoref{fig:entity_rela}).
    \item[(B)] If the query contains only one entity, then the description of the entity is used for query expansion.
    \item[(C)] If the query contains more than one entity, but it does not contain any relationship, then only entities are used for query expansion.
\end{enumerate}
% (A) is ideal, with all information we need to find potential information in the external knowledge. However, in most of the cases, the query contains only entities, without any relationships. 
In case (A), one can navigate the KG to find relevant information on the KG for the query. However, many queries do not contain any relations for KG navigation. To still benefit from KG in such cases, we use entities information on the KG for case (B) and (C).
 
The expanded query is then (1) used in \emph{candidate generation} to rank the candidates and (2) used  in \emph{explanation generation} to highlight the \mis{}.

% Entity relationship diagram

%\begin{example}
% \includegraphics[scale=0.2]{figures/entity_rela.pdf}
\begin{example}
\normalfont
\autoref{fig:entity_rela} exemplifies entity linking and relationship extraction used to identify the \mis{}. The query is \textit{cause of heart disease}, and the extracted entity and relationship are \textit{heart disease} and has \textit{contributing factor}. So the linked (unique) entities are obesity and atherosclerosis which are thus used for query expansion.
%Each entity and relationship has a unique id that starts with Q and P.
\end{example}

%\end{example}

%\subsubsection{Visualization of the relationship between query and retrieved keyword}
%The second explainable element is identifying the relationship of query and keyword of the retrieved passage using a \kg{}. 

%\paragraph{Rational}
%Viewing the relationship between query, link, and retrieved keyword can give users a clear understand of why this passage is retrieved and what is the main content of the passage. 
% example???

%\paragraph{Explanation with Knowledge Graph}
%The initial steps are the same as visualization of the \mis{}, including entity recognition, relationship extraction, and entity linking with knowledge graph. After identifying the entity and links within the query, we can look into the entity's properties or relationships to see if the answer is contained by the related entity.

%'Glacier caved' 'formed' 'glacier'

% example???

\subsection{\kg{} for Explainable Re-ranking}
\label{sec:ex-re-ranking}
% Another explainable element is  re-ranking of documents with explainable metrics extracted from knowledge graph.
%Compared with vector-based approaches, linear ML models, such as logistic regression provides intrinsic explainability based on feature importance. 
%By re-ranking documents with a linear combination of explainable metrics, the system can better answer the question of why a document is ranked at a certain position. 

User studies has shown that re-ranking based on \emph{explainable features} can improve the interpretability of the ranking result \cite{polley2021exdocs} since users can get insights into the passage by observing only the feature values. Examples of such explainable features include term statistics, contextual words, and citation-based popularity \cite{polley2021exdocs}.

%can better answer the question: of why a document is ranked at a position. 
%Moreover, users can also get insights into the passage by only looking at explainable metrics. %without reading through the whole passage.
%It was shown in 

% \paragraph{Explainable Metrics from \kg{}}
Our framework incorporates the idea of re-ranking documents along explainable features to better explain the ranking of a document.
We propose a new explainable feature used for re-ranking candidate documents based on the \kg{} with entity matching. 
%One explainable metric that can be extracted from \kg{} is entity matching. 

Entity matching finds how two text passages are related to each other by (1) extracting entities from two passages \cite{vanHulst:2020:REL}, and (2) calculating the pair-to-pair relatedness between two types of entities \cite{tagme}. The authors of \cite{tagme} present a measure of relatedness $R$ based on the overlap of the incoming links between two entities on graphs such as Wikipedia KG. Given a pair of phrase, the algorithm will find the corresponding entities on the KG $P_a$ and $P_b$ and then compute the relatedness between $P_a$ and $P_b$ as suggested in \cite{tagme} as follow:
\begin{equation}
    R(P_a, P_b) = \frac {log(max(|in(P_a),in(P_b)|)) - log(|in(P_a) \cap in(P_b)|)}{log(W) - log(min(|in(P_a), in(P_b)|))}
    \label{equ:single_relatedness}
\end{equation}
where $in(P_a)$ is the set of incoming edges to entity
$P_a$ and W is the total number of nodes in the KG.

To get an entity relatedness score between the query and a document, we average the relatedness values along document entities and then sum up along query entities. Specifically, the \emph{query-document relatedness} score is defined as follows:
\begin{equation}
    QDR = \sum_{i=0}^{i=n} \frac {\sum_{j=0}^{j=m} R(E_{q_i}, E_{d_j})}{m}
    \label{equ:relatedness}
\end{equation}
% Formula to calculate the query-document relatedness
where $R$ is the relatedness score function \cite{tagme} between two entities, $E_q$ is a set of $n$ entities from the query, and $E_d$ is a set of $m$ entities from the document.
This value could be used as an explainable feature to re-rank documents proposed by \textit{candidate generation}.

Though this approach ignores structure from the sentence such as the order of entities. the computation can be done efficiently in parallel for each individual entity. We leave exploring structure information of the sentence as future work.

%- Entity Matching \\
%- Opt1: Find most relevant passage using the computed feature
%- Opt2: Compute Reranking Features from KG \\ (Learning on Graph)

% 1. Input/output for each step
% 2. Describe each step of the process
% \clearpage
\section{Evaluation}
\label{evaluation}

In this section, we investigate the following research questions: 
\begin{itemize}
%  \item RQ1: How knowledge graph would affect the performance of the information retrieval?
 \item RQ1: How can a \kg{} influence the performance of highlighting the most important sentence?
\item RQ2: How can a \kg{} affect the performance of retrieval by re-ranking with explainable features?
\end{itemize}

\subsection{Experiment Setup}
% In this section, we provide a list of datasets and \kg{}s we used in the experiments. We selected Wikidata as our main  reference \kg{}. 

\paragraph{Reference \kg{}}
Wikidata \cite{Wikidata} is a open knowledge base containing structured data with 99,959,412 data items from multiple domains. We select Wikidata as the reference \kg{} as it is often used for tasks such as link prediction \cite{Galkin2022AnOC}, knowledge graph retrieval \cite{waagmeester2020science} and recommendation \cite{alghamdi2021learning}.
%At the same time, The data is in RDF format and can be queried via SPARQL. 
In the experiment, we first use the entity linker to identify entities and relations from the query and then navigate the \kg{} from the matched entities and relations (using the \textit{qwikidata} package \cite{qwikidata}). At the same time, we also used Wikipedia page graph \cite{wiki} to identify the relatedness of two entities.  
% github website: https://github.com/kensho-technologies/qwikidata
%there are many other well-developed packages for accessing Wikidata.

\paragraph{Evaluation datasets}
To evaluate the performance of \emph{identifying the \mis{}}, we select WIKIQA \cite{YangYihMeek:EMNLP2015:WikiQA}, which is a popular multi-domain question-answering dataset and benchmark used by the \ir{} community. 
%WIKIQA \cite{} is a public open-domain question-answering dataset where 
%In WIKIQA, questions originated from Bing query logs, and documents are created from Wikipedia page summaries. 
Unlike most other \ir{} datasets,
%this dataset contains 3,047 questions and 29,258 sentences, 
sentences in this dataset are labeled as answers to their related questions. We choose this dataset for two main reasons: (1) high quality answers are selected from the Wikipedia page summary, and (2) the sentences that answer the question are labeled for evaluating the performance of sentence-level retrieval. %We used WIKIQA distribution from HuggingFace datasets \cite{YangYihMeek:EMNLP2015:WikiQA}. 

% The primary reason why we used a QA dataset in evaluating this task instead of traditional information retrieval datasets like TREC\cite{Text_retrieval_conference} is that many QA datasets only label for the best document and identifying sentences correctly answering the question.   

To evaluate the performance of \emph{re-ranking documents with explainable metrics}, we use the test set from TREC 2004 Robust (Robust04) \cite{robust04}, which is  popular benchmark to evaluate the performance of text-based \ir. Robust04 contains 249 queries and 528,000 corpora from the news. Each query is associated with multiple relevant documents for document ranking evaluation. Moreover, this dataset assigns various levels of document relevance to a query, which makes it suitable to assess the ranking performance.

%is complex enough so that it gives room for potential improvement.

% compensate some shortcomings from WIKIQA. Firstly, WIKIQA has only 1 or 0 correct document for one query so the situation of retrieving and ranking multiple relevant documents would not happen within this dataset. Secondly, WIKIQA is a small dataset and the the passage-level retrieval performance has already been very high with baseline approach and this is no room for possible improvement. However, TREC 2004 Robust perfectly resolved the above-mentioned 2 issues. It has multiple relevant documents for each query and the baseline approach performed poorly on this dataset.  

% description
% Statistics
% Why this dataset (has \mis{} label)

% Description
% Statistics
% Why this dataset (has \mis{} label)

\paragraph{Compared approaches}
% Baseline: No Knowledge Graph
To demonstrate the benefit of extra knowledge from \kg{}, we rely on a pre-trained embedding retriever with a pre-trained sentence transformer model \cite{reimers-2019-sentence-bert, flax-sentence-embeddings_2021} for all compared approaches.
% \textit{all-mpnet-base-v1} 
%to retrieve the \mis{}. , we keep same embedding model throughout the experiments. 
We use Haystack \cite{haystack} to build the \ir{} pipeline for all compared approaches and use FAISS \cite{johnson2019billion} to perform quick vector ranking based on cosine similarity. 

In the case of \emph{identifying \mis{}}, we first split the retrieved documents into individual sentences.
% new retriever with the same embedding model will retrieve sentences among this temporary document store with the same query/question. 
A retriever with the pre-trained embedding model will rank the sentences using cosine similarity with the query. Finally, the sentence with the highest similarity score will be treated as the \mis{} (see \autoref{equ:mis}), which will be compared with the ground-truth sentence to check the performance. Unlike the baseline (with no query expansion), our approach expands the query with information from the \kg{}. 
%We keep the retriever the same and only enrich the query used to find the \mis{}.

In case of \emph{explainable re-ranking}, the baseline approach also apply the embedding retrievers to rank documents with cosine similarity. The baseline directly uses the ranking from the embedding retriever and do not re-rank the documents. In our approach, we use the same retriever to generate the candidates but re-rank them using \emph{query-document relatedness} from \autoref{sec:ex-re-ranking}.

\paragraph{Evaluation metrics}
To evaluate the performance of the approaches, we use several metrics widely adapted by the \ir{} community. Specifically, \textbf{Accuracy} is used to evaluate the performance of identifying \mis{}. To assess the performance of re-ranking,  we use Precision (\textbf{P}) and  \textbf{Recall} to measure the quality of retrieved documents. However, since these metrics do not consider the ranking, we use Mean Average Precision at k (\textbf{MAP@k}) and Normalized Discounted Cumulative Gain at k (\textbf{NDCG@k}) to evaluate the quality of ranking for the top \textit{k} documents.

%\subsection{Research Questions}

\subsection{RQ1: MIS as Explanation}
\paragraph{Rationale}
In this RQ, we aim to evaluate the influence of extra knowledge incorporated into the query from the \kg{} on the performance of highlighting the \mis{}. 
%\todo{This is new}
% We use KGs for query expansion and compare the performance of highlighting the \mis{} with the baseline approach. 
\paragraph{Setup}
% Entity linking, knowledge graph, dataset, 
% How the measurement is conducted
We use the query from the WIKIQA dataset and the Wikidata knowledge base for query expansion. In the baseline approach, we use the original query to retrieve the most relevant passage and then retrieve the \mis{} within the passage. Our approach performs the same retrieval process using the expanded query. For the evaluation, we calculate the accuracy of retrieving the top \mis{} for both approaches. We measure the performance for identifying \mis{} using the test set with 243 queries and 1081 documents. 
%\todo{fill the number here}

We measure the performance for two variations of our approach in query expansion: (1) using an existing entity matching tool (\textbf{ours 1}) and (2) using an ideal entity matching approach (\textbf{ours 2}). We use the same reference KG in both cases. In \textbf{ours 1}, we used WAT \cite{wat} to identify entities within the query and link to Wikidata items. However, this automated process can be erroneous and may not reflect the benefit of extra information from the \kg{}. 
To illustrate the potential benefit of \kg{}s, in \textbf{ours 2}, we manually reviewed the results of WAT, identified entities with relationships, and linked them to the correct entities. In this approach, we assume an ideal performance of the entity matcher for query expansion and test the performance of sentence retrieval.

\paragraph{Discussion}
% Result with WAT
% Result with WAT + manual check
% Here are the main difference and performance for three approaches:
% \begin{itemize}
% \item \textbf{baseline approach}: original query from WIKIQA dataset
% \item \textbf{approach 1}: use WAT for entity recognition and entity linking. Use the entity description from Wikidata for query expansion.
% \item \textbf{approach 2}: check the entity linking performance of WAT, label the entity and relationship manually, and use the information from Wikidata for query expansion.
% \end{itemize}

\begin{table}[t]
    \centering
     \begin{tabular}{|c|c|c|}
    %{0.45\textwidth}
    % { 
    %   | >{\raggedright\arraybackslash}X 
    %   | >{\centering\arraybackslash}X 
    %   | >{\raggedleft\arraybackslash}X | }
     \hline
      & Passage Retriever & Sentence Retriever \\
     \hline
     baseline  & 97.53 & 66.67  \\
      \hline
     ours 1  & 97.12 & 62.71  \\
     \hline
     ours 2 & \textbf{97.94}  & \textbf{71.31}  \\
    \hline
    \end{tabular}
    \caption{Accuracy for sentence retrieval (in \%)}
    \label{tab:highlight}
\end{table}
 In \autoref{tab:highlight}, the \textit{ passage retriever} column illustrates the accuracy of the original \ir{} task of finding the most relevant document to the query. The \textit{sentence retriever} shows the accuracy of identifying \mis{}.
 
The accuracy of the passage retriever is high for all three approaches (more than $98.00\%$). The accuracy of \textbf{ours 1} is slightly lower than the baseline approach by $0.77\%$, and the accuracy of \textbf{ours 2} is higher than the baseline approach by $0.24\%$. The performance for the sentence retriever shows a similar trend. While the accuracy of \textbf{ours 1} is lower than the baseline by $1.79\%$, the result of \textbf{ours 1} outperforms the baseline by $7.29\%$.

After an in-depth investigation of query expansion, we found that the external entity matching tool used in \textbf{ours 1} was not accurate for some queries. In this case, the query is expanded with misleading information and does not select the best \mis{}, which causes performance decrease. However, with an ideal entity matcher (\textbf{ours 2}), the performance of identifying \mis{} is improved significantly. 

% \paragraph{Conclusion}
\ranswer{1}{We find that the performance gain for highlighting the \mis{} is dependent on the performance of the entity matching and the relevance of the \kg{}. Using an ideal entity matcher, the additional information from the \kg{} significantly improves the accuracy of sentence retrieval without modifying the retrieval model. }

% Answer the research question
% We found the performance of highlighting the \mis{} is not good in the early investigations due to the poor performance of entity linking. However, in the second experiment, we found the accuracy of finding the \mis{} is better than the baseline approach. So we suspect the performance of query expansion and retrieving the \mis{} depends on the quality of entity linking and relevancy of the KG.

\subsection{RQ2: Explainable Re-ranking}
\paragraph{Rationale}
Entity matching and query-document relatedness are used as explainable features to re-rank documents, and the re-ranking result is then compared with the baseline approach. While these features can improve the explainabilty of the ranking, we aim to evaluate their influence on the ranking performance. 
% \todo{this is new}

\paragraph{Setup}
The Robust04 dataset was used in this part of the evaluation. In the baseline approach, an embedding retriever is used to retrieve and rank documents based on cosine similarity with original queries and documents from the dataset. Our approach (\textbf{ours}) retrieves documents with the same embedding retriever, but the documents are ranked based on the query-document relatedness score (see \autoref{equ:relatedness}). Specifically, it uses REL \cite{vanHulst:2020:REL} for entity linking and TagMe \cite{tagme} for calculating entity relatedness scores and combines these two techniques to compute query-document entity relatedness scores for ranking documents. The quality of the ranked documents is measured by the TERC evaluation tool \cite{VanGysel2018pytreceval}.
% After getting two lists of ranked documents, they were sent to the evaluation tool to evaluate the performance of ranking. 

\paragraph{Discussion}
As shown in the \autoref{tab:reranking}, both methods have the same \textbf{P} and \textbf{Recall} because re-ranking does not exclude existing candidates from candidate generation. Moreover, these two metrics do not measure the ranking performance. While our approach improves explainability of results (thanks to the metrics derived from the \kg{}), it achieves 8.13\% \textbf{MAP@20} and 17.89\% \textbf{NDCG@20} which are slightly lower than the ones of the baseline approach (by 2.25\% and 2.98\%).

To further investigate this drop in performance, we compared the result ranking of the two approaches and found that only 17.95\% queries have better ranking results with the new approach while ranking stays the same for the majority of the queries. After investigating entity linking for queries, we noticed  similarly limited performance of entity linking as in \textit{RQ1}. This finding suggests that a better entity linking may also potentially improve the performance of the re-ranking performance. 

% \paragraph{Conclusion}
\ranswer{2}{We find that while explainable features from the \kg{} increase the explainability of the ranking, they may decrease the retrieval performance. We suspect this decrease is rooted in imperfections of the entity linker. Additionally, our finding is in line with others \cite{polley2021exdocs} that explainable re-ranking decreases the ranking performance.}

\begin{table}[tb]
    \centering
     \begin{tabular}{|c|c|c|c|c|}
    %{0.45\textwidth}
    % { 
    %   | >{\raggedright\arraybackslash}X 
    %   | >{\centering\arraybackslash}X 
    %   | >{\raggedleft\arraybackslash}X | }
     \hline
      & P & Recall & MAP@20 & NDCG@20 \\
     \hline
     baseline  & 31.09 & 15.71 & 10.38 & 20.87  \\
      \hline
     ours  & 31.09 & 15.71 & 8.13 & 17.89 \\
    \hline
    \end{tabular}
    \caption{Evaluation result for re-ranking (in \%)}
    \label{tab:reranking}
\end{table}

\subsection{Threats to Validity}
\paragraph{Internal validity}
The training of embedding models is normally non-deterministic. In order to minimize the effect of randomness, we used a pre-trained embedding model \cite{reimers-2019-sentence-bert, flax-sentence-embeddings_2021} and kept it the same throughout the experiment. 

\paragraph{External validity}
We used Wikidata as our reference \kg{} throughout the experiments. Obviously, one may get different improvements by using \kg{} from a different domain. In the experiment of \textit{RQ1}, we measured performance using an existing entity matchers \cite{wat, vanHulst:2020:REL} and assuming an ideal entity matcher. The real-world performance may be in between the two. Similarly, in the experiment of \textit{RQ2}, we re-ranked the documents with entity relatedness using the \kg{}, and one may get a different result by incorporating different explainable features using the \kg{}.

\paragraph{Construct validity}
We measured the performance of compared approaches using popular \ir{} benchmarks on classic \ir{} metrics, which follows the best practices in \ir{} evaluations.
% \clearpage
%\documentclass[../main.tex]{subfiles}
%\begin{document}
% ~1 page Wednesday
\section{Related Work}
\label{sec:rel-work}
\subsection{Explainable Artificial Intelligence}
 Generally, XAI approaches can be divided into two categories: \emph{ante-hoc} methods 
%  (involved in the training stage) 
 and \emph{post-hoc} methods (by using an external explainer on an already trained model) \cite{islam2022systematic, vilone2020}. The latter can be further divided into two categories, \emph{model-specific} and \emph{model-agnostic}. 
 
\textit{Ante-hoc methods} create transparent models so that the explanation can be generated directly from the models, such as tree-based models \cite{sagi2020explainable, itani2020one} or extending the architecture of neural networks for explanation \cite{rio2020understanding, sarkar2022framework, csiszar2020interpretable}. There are also many surveys on post-hoc methods for explanation \cite{zhou2021evaluating, verma2020counterfactual, islam2022systematic}. 

\emph{Model-specific} methods are confined to particular models by getting explanations using the internal model representation and learning process, and can generate explanation directly from model output. However, most of the methods only work with simple interpretable models such as linear models, nearest neighbours and tree-based models \cite{islam2022systematic}. 
% Wagner et al. \cite{wagner2019interpretable} and wang et al. \cite{wang2020score} create local visual explaination for CNNs based on the gradient.  
% Model-specific methods usually generate explainable result with intrinsically interpretable models such as linear model, nearest neighbour model, and tree-based model 
%Model-specific methods can generate explanation directly from model output. However, most of the methods only work with simple interpretable models such as linear models, nearest neighbours and tree-based models.. 
Instead, \emph{model-agnostic} methods only analyze input and output pairs to improve explainability without looking into the model internals. 
 %Popular model-agnositic methods to improve explainability include 
 %LIME and SHAP \cite{lundberg2017unified}, which generally have good performance and they are widely used in various applications. 
 One popular model-agnostic methods, 
 %\emph{Local Interpretable Model-Agnostic Explanations} 
 \textit{LIME} \cite{zhang2019should} was proposed to explain the predictions of any classifier in an interpretable and faithful manner.
 %by training an inherently interpretable model on a new dataset made from the permutation of samples and the corresponding prediction of the black box model. However, the learned model does not have a good global approximation even if it performs well on a local level. Moreover, it sometimes provides very different explanations for two nearby datapoints  \cite{islam2021explainable}. 
 Another popular method, \textit{SHAP} \cite{lundberg2017unified} is built based on the Shapley regression values, which is an approach to compute feature importance and allow fast approximations in situations where training models on all feature subsets would be intractable. 
 %However, it also suffers from the problem of correlated features. Furthermore, the Shapely value returns a single value per feature; there is no way to make a statement about the changes in output resulting from the changes in input \cite{islam2021explainable}.
 However, these methods suffer from problems such as poor global explanation and correlated features \cite{islam2021explainable}.
 
 In this paper, we proposed a general framework to generate \textit{model-specific} \textit{post-hoc} explanation for embedding-based information retrieval models.

\subsection{Explainable Information Retrieval}
Explainability in information retrieval can be categorized by the form of the outcome: highlighting-based, feature importance based, rules-based, and mixed approaches \cite{islam2022systematic}. 
% The expl the IR system is keyword-based or embedding-based.  \todo{types of output}

One famous example for \emph{highlighting-based} explanation is Google Search, where content snippets and keywords are shown along with the search result. Similarly Chios and Verberne \cite{p2} proposed highlighting most important snippet for deep IR models.
%Many information retrieval applications offer mixed formed of explainable elements. Examples can be found from Google search. The result of Google search will show not only the title-bar of the website but also a snippet of the website content and highlighted keywords in the snippet, which already includes visualization and text. These features integrate colors and shapes to make search results explainable to users. 
%Chios et al. \cite{p2} also proposed a similar feature of highlighting most important snippet for deep IR models. 
\emph{Feature importance based} explanation methods may calculate scores from queries and documents \cite{Ramos2019ExplainabilityIT, p2, polley2021exdocs}. Some approaches operate on term importance from the query, while others focus on re-ranking the documents with explainable features \cite{polley2021exdocs}.
%. The way of calculating term importance varies. While some approaches operate on term importance from the query, others focus on reranking the documents with explainable features \cite{polley2021exdocs}.
%While \cite{Ramos2019ExplainabilityIT} derive term importance from BM25, \cite{p2} calculated term importance by using  term gating network’s softmax function.
\emph{Rule-based explanation} aims to answer specific properties about the internal mechanism of the IR model \cite{p3} by providing a simplified model to explain decisions.
% Some papers also attempted to generate rules to explain the internal mechanism of information retrieval model \cite{p3}.
Finally, \emph{mixed} explanation combines all methods to provide comprehensive explainability \cite{p2,polley2021exdocs}.

While explainable \ir{} is mostly post-hoc, there also exist approaches using explainable embedding to rank documents \cite{qureshi2019eve, panigrahi2019word2sense} such that the distance in the vector space explains the relatedness of the query and document.  
% For example, \cite{p3} is looking into the interpretability of neural retrieval models to get more insight on how search algorithm works, which modified LIME \cite{zhang2019should} to reveal mechanism of neural ranking by showing the contribution of terms in the document. 

Compared with these approaches, our general framework generates mixed forms of explanations for IR systems with an external knowledge graph. Furthermore, we propose a concrete instantiating of the framework to highlight the most important sentence and explainable re-ranking.

\subsection{Knowledge Graph for Information Retrieval}
Entities from a knowledge graph can be used within an \ir{} system in order to help understand of a user’s intent, queries, and documents beyond what can be achieved through word tokens on their own \cite{p5}. Reinanda et al. \cite{p5} summarised approaches on how to leverage entity-oriented information in \kg{}: \emph{expansion-based}, \emph{latent factor modeling}, \emph{language modeling}, and \emph{deep learning approaches}. At the same time, KG is also being used in some domain-specific applications for explanation of IR systems. 

\emph{Expansion-based} approaches enrich entity-oriented information in the retrieval process by expanding queries and/or documents \cite{balaneshinkordan2016empirical}, for example, expanding the query with synonyms.

\emph{Latent factor modeling} attempts to find concepts in queries and documents. The authors of \cite{xiong2015esdrank} presented a new technique for improving ranking using external data and knowledge bases. This technique treats the external objects as latent layer between query and documents to learn judging document relevance.

\emph{Language modeling} approaches consider semantic information when building language models of queries and documents. For example, Ensan et al. \cite{ensan2017document} proposed a document retriever which uses semantic linking systems for forming a graph representation of documents and queries, where nodes represent concepts from documents and edges represent semantic relatedness between concepts. 

\emph{Deep learning} approaches uses knowledge graph embeddings in neural ranking systems. Examples are embedding queries and documents in the entity space, \cite{xiong2017explicit} and constructing an interaction matrix between queries and entity representations. \cite{liu2018entity, xiong2017word} 

\emph{KG for XIR}
Hasan et al \cite{abu2022transferrable} proposed a  framework for generating explanation in domain-specific IR applications by incorporating domain knowledge graphs. The author of \cite{yang2020biomedical} applies \kg{} to explain the \ir{} model by building knowledge-aware paths with the help of attention scores. Similarly, the author of \cite{xian2019reinforcement} designs an explainable recommendation system which contains explicit reasoning with \kg{} for decision making to make recommendations explainable.

%Different from these approaches, 
In this work, we proposed a general framework for utilizing knowledge graphs to improve the \textit{explainability} in IR systems. 

\section{Conclusions and Future Work}
\label{conclusion}
In this paper, we investigate the task of XIR, targeting to explain why a document is relevant to a query. While existing approaches use highlighting and feature importance, less focus has been put on using an external knowledge base to generate explanations. 

We propose a general architecture that uses semantic information from \kg{}s to improve the explainability of \ir{} systems. We take advantage of existing entity and relationship matching methods to identify the most relevant part of the \kg{} to a passage and navigate the \kg{} to help explain the result of retrieval from each process in \ir{}. We demonstrate the effectiveness of \kg{}s with two examples \ir{}: highlighting the most important sentence and re-ranking with explainable features. 

We carried out an initial experimental evaluation of our approach using multiple metrics with the key finding that the performance of our approach largely depends on the quality of entity matching. With an erroneous entity linker, the performance can decrease compared to the baselines.  However, with an ideal entity matcher, our technique improved performance of identifying \mis{} by $6.96\%$ and the base retrieval performance by $0.45\%$.

We believe this general architecture opens many directions for using \kg{} to improve the explainability of \ir{}. In the future, we aim to apply the general architecture to  (1) combine the architecture with a more advanced entity/relation matching methods, (2) adapt various knowledge graph retrieval methods to improve the explainability of \textit{candidate generation}, (3) integrate \kg{} with the embedding model to create explainable text embeddings, (4) capture structure information from documents and queries and (5) evaluate how our approaches influence the explainability of IR system in human's perspective with user study. The code and artifacts for the experiments can be found in \footnote{\url{https://github.com/Aggregate-Intellect/xir}}

%In this paper, we proposed two approaches to improve the explainability of IR system using external knowledge base, highlighting the most important sentence and re-ranking with explainable features. 
% We also evaluated two approaches using accuracy, precision, recall, MAP@20, and NDCG@20. In the second approach of KG for Explainable Re-ranking, one defect is that the algorithm ignores structure information such as word orders and phrases. We will attempt to figure out a better way to capture structure information from documents and queries in our future work. Due to lack of ground truth explanation, we only focused on evaluating the explanation generation from the retrieval point of view. In the future, we may also evaluate how these approaches influence the explainability of IR system in human's perspective such as user study.

\section{Acknowledgement}
This paper is partially supported by the NSERC RGPIN-2022-04357 project and the FRQNT-B2X project (file number: 319955). This work is done as part of the McGill Summer Undergraduate Research in Engineering 2022 project (ECSE-025).
% \input{sections/Future-Work}

%In this paper, we approach to consider the constraints for scene graph generation explicitly as a post processing process to a neural network-based scene graph generator. 

%%
%% The next two lines define the bibliography style to be used, and
%% the bibliography file.
\bibliographystyle{ACM-Reference-Format}
\bibliography{main}

%%
%% If your work has an appendix, this is the place to put it.
\clearpage
% \appendix
% \input{sections/appendix}
\end{document}